\documentclass[%
reprint,
 amsmath,amssymb,
aps,
prb,
groupedaddress,
]{revtex4-1}
\usepackage{graphicx}
\usepackage{dcolumn}
\usepackage{bm}

\begin{document}

\preprint{APS/123-QED}

\title{YIG thickness and frequency dependence of the spin-charge current conversion in YIG/Pt systems}

\author{V. Castel}
\affiliation{ 
University of Groningen, Physics of nanodevices, Zernike Institute for Advanced Materials, Nijenborgh 4, 9747 AG Groningen, The Netherlands.
}%

\author{N. Vlietstra}
\affiliation{ 
University of Groningen, Physics of nanodevices, Zernike Institute for Advanced Materials, Nijenborgh 4, 9747 AG Groningen, The Netherlands.
}%

\author{J. Ben Youssef}%
\affiliation{ 
Universit\'e de Bretagne Occidentale, Laboratoire de Magn\'etisme de Bretagne CNRS, 6 Avenue Le Gorgeu, 29285 Brest, France.
}%

\author{B. J. van Wees}
\affiliation{ 
University of Groningen, Physics of nanodevices, Zernike Institute for Advanced Materials, Nijenborgh 4, 9747 AG Groningen, The Netherlands.
}%

\date{\today}

\begin{abstract}

We report the frequency dependence of the spin current emission in a hybrid ferrimagnetic insulator/normal metal system as function of the insulating layer thickness. The system is based on a yttrium iron garnet (YIG) film [0.2, 1, and 3 $\mu$m] grown by liquid-phase-epitaxy coupled with a spin current detector of platinum [6 nm]. A strong YIG thickness dependence of the efficiency of the spin pumping has been observed. The highest conversion factor $\Delta\textrm{V}/P_{\texttt{abs}}$ has been demonstrated for the thinner YIG (1.79 and 0.55 mV/mW$^{-1}$ at 2.5 and 10 GHz, respectively) which presents an interest for the realisation of YIG-based devices. A strong YIG thickness dependence of the efficiency of the spin pumping has been also observed and we demonstrate the threshold frequency dependence of the three-magnon splitting process.

%
\end{abstract}

\maketitle

Recently in the field of spintronics, Y. Kajiwara \textit{et al.}\citep{Kajiwara2010nature} opened a renewed interest by the demonstration of the spin pumping and inverse spin Hall effect (ISHE) processes in a hybrid system based on yttrium iron garnet (YIG) coupled with a layer of platinum (Pt). The YIG/Pt system presents an important role for future electronic devices based on non-linear dynamics effects\citep{SaitohBistable,Kurebayashi2011nmat, HillebrandsSPmagnons, DemoParametric,2mag}, such as the three-magnon splitting process. The three-magnon splitting process presents a frequency (or magnetic field) dependence and a YIG thickness dependence\citep{YIGPt01,Chernyshev} which permits to tune the conversion efficiency of a spin current from spin pumping as function of these parameters.

In this paper, we show the experimental observation of the YIG thickness dependence in a hybrid YIG/Pt system of the dc voltage generation from spin pumping, actuated at the resonant condition over a large frequency range. We demonstrate that the three-magnon splitting process ceases to exist for the thinner YIG of 0.2 $\mu$m, which is way smaller than the exchange interaction length in such system.

\begin{figure}
\centering
\includegraphics[width=8.5cm]{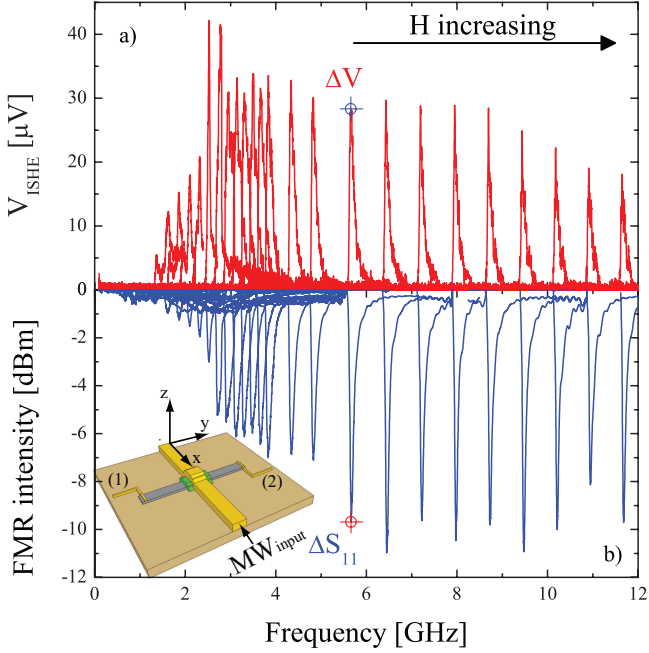}
\caption{\label{fig:Fig1} 
Frequency dependence of: a) the dc voltage from spin pumping $V_{\textrm{ISHE}}$, b) the FMR response determined by the scattering parameter $S_{11}$ for different values of the static magnetic field ($H$) applied in the plane of the bilayer along $x$. The thickness of the YIG layer is 1 $\mu$m. For each value of $H$, the frequency excitation has been swept at an rf power of 10 mW (at room temperature). $\Delta\textrm{V}$ and $\Delta S_{11}$ correspond to the magnitude of $V_{\textrm{ISHE}}$ and $S_{11}$ at the resonant condition, respectively. The FMR response results from the difference between two frequency sweeps, one at the resonant magnetic field ($H_{\textrm{res}}$) and the other one at a saturation value ($H_{\textrm{sat}}=4$ kOe). The inset represented the measurement setup configuration. (1) and (2) are the electrical contacts, the gray area corresponds to the Pt layer, the green to the Al$_2$O$_3$ layer, and the brown part to the YIG.
}
\end{figure}

The used insulating material consists of a single-crystal (111) Y$_3$Fe$_5$O$_{12}$ (YIG) film grown on a (111) Gd$_3$Ga$_5$O$_{12}$ (GGG) substrate by liquid-phase-epitaxy. Three samples with different thicknesses of YIG [0.2, 1, and 3 $\mu$m] have been prepared. For each sample, a spin current detector based on a platinum (Pt) layer of 6 nm grown by dc sputtering has been used\citep{YIGPt02}. Schematic of the microwave measurement setup is shown in the inset of Fig.\ref{fig:Fig1} b). Several steps of electron beam lithography have been done in order to pattern the Pt area ($600\times 30 \mu$m), the insulating layer of Al$_2$O$_3$ between the Pt layer and the stripe antenna (SA), and finally a Ti/Au deposition for the SA (60 $\mu$m width) for the rf excitation and electrodes for the electrical contacts. A signal-ground picoprobe has been used in order to connect the SA to the network analyser.


The measurement setup for this investigation is different from Ref.\citep{YIGPt01,YIGPt02}. In the present case, a simultaneous detection of the dc voltage generation (without modulation and lock-in) in the Pt layer ($V_{\textrm{ISHE}}$) and of the ferromagnetic response (FMR intensity) has been performed, as shown in Fig.\ref{fig:Fig1} a) and b), respectively. In these figures, $\Delta\textrm{V}$ and $\Delta S_{11}$ correspond to the magnitude of $V_{\textrm{ISHE}}$ and to the microwave absorption power at the resonant condition $f_{\texttt{FMR}} $, respectively. $S_{11}$ corresponds  to the reflection coefficient extracted from the scattering parameter of the network analyser (in one port configuration). $V_{\textrm{ISHE}}$ comes from the fact that at the resonant condition ($ f_{\texttt{FMR}} $) a spin current ($j_{s}$) is pumped into the Pt layer and converted in a dc voltage due to the ISHE. In this system, the pumped spin current originates from the spin exchange interaction at the interface between localized moments in YIG and conduction electrons in the Pt layer. The static magnetic field, $H$, is applied in the plane of the device and oriented perpendicularly to the length of the Pt layer (along $x$, see inset Fig.\ref{fig:Fig1} b)). In this configuration, the signal is maximum and the sign of $V_{\textrm{ISHE}}$ is changed by reversing the magnetic field along $-x$ (not shown). One can see in Fig.\ref{fig:Fig1} b) that the FMR line is broadened due to the contribution of the Backward Volume Magnetostatic Spin Waves (BVMSW, when $f<f_{\texttt{FMR}} $) and from the Magnetostatic Surface Spin waves (MSSW, when $f>f_{\texttt{FMR}} $)\citep{SWstancil,HilleSelection}.

\begin{figure}
\centering
\includegraphics[width=9.3cm]{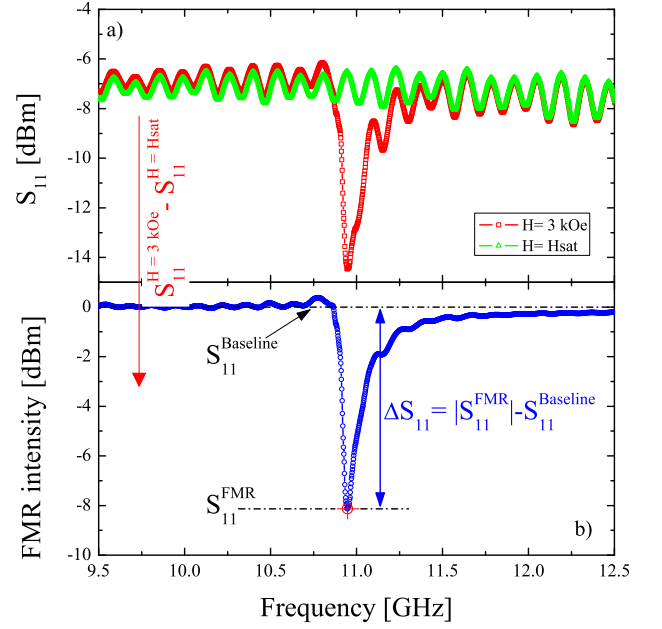}
\caption{\label{fig:Fig2bis} 
a) Frequency dependence of $S_{11}$ spectrum measured at H = 3 kOe (in red) and at a saturation field, $H_{\textrm{sat}}$ (in green) equal to 4 kOe for a YIG thickness of 1 $\mu$m. The measurement has been performed at an rf power of 10 mW at room temperature. b) The frequency dependence of the FMR intensity corresponds to the difference between the $S_{11}$ spectrum for the resonance in the YIG layer (H = 3 kOe) and from the spectrum without resonance signature ($H_{\textrm{sat}}$). $S^{\texttt{Baseline}}_{11}$ and $S^{\texttt{FMR}}_{11}$ correspond to the magnitude in dBm of the baseline and at the resonant condition FMR.
}
\end{figure}

The FMR intensity in Fig.\ref{fig:Fig1} b)  corresponds to the difference between the $S_{11}$ spectrum for the resonance in the YIG layer and from the spectrum without resonance signature. An example is presented in Fig.\ref{fig:Fig2bis} for a static magnetic field of 3 kOe. In order to define the FMR absorption $P_{\texttt{abs}}$ in mW, the following equation has been used: $P_{\texttt{abs}} [\texttt{mW}]\propto 10^{\mid S^{\texttt{FMR}}_{11}\mid /10} - 1$. The magnitude of the baseline in dBm for the set of sample and for different values of the static magnetic field is equal to 0 dBm (1 mW).

\begin{figure}
\centering
\includegraphics[width=9.3cm]{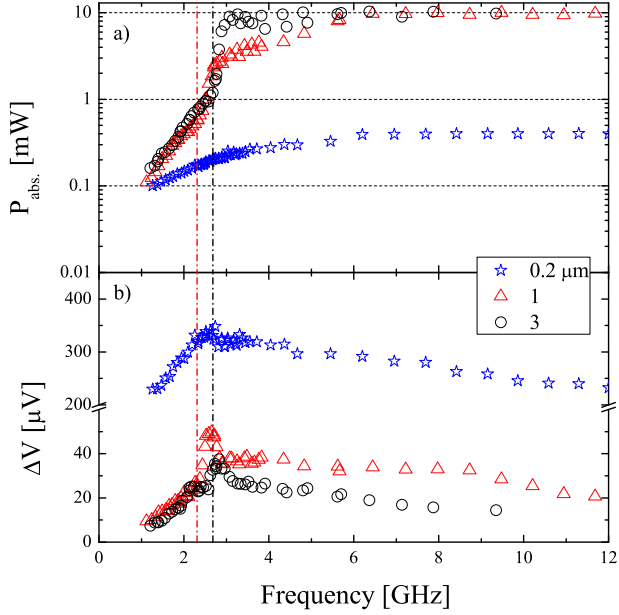}
\caption{\label{fig:Fig2} 
Frequency dependence of: a) the FMR absorption $\Delta S_{11}$ converted in mW and b) the dc voltage $\Delta\textrm{V}$ for different thicknesses of YIG (0.2, 1, and 3 $\mu$m). Vertical dash lines correspond to the frequency cutoff of the three-magnon splitting process estimated with the dispersion relation of spin waves for a YIG thickness of 1 (red) and 3 $\mu$m (black). All Measurements have been carried at room temperature under an rf excitation of 10 mW.
}
\end{figure}

Similar measurements as those presented in Fig.\ref{fig:Fig1}  have been performed for the different thicknesses of YIG. For each sample, the dc voltage generation ($V_{\textrm{ISHE}}$) and of the ferromagnetic response ($S_{11}$) have been studied as function of the frequency and the applied magnetic field at 10 mW. The magnitude of theses dependences ($V_{\textrm{ISHE}}$ and $S_{11}$) at the resonant condition have been extracted for the set of sample. Fig.\ref{fig:Fig2} a) and b) give a summary of the frequency dependence of the FMR absorption $P_{\texttt{abs}}$ and of the dc voltage $\Delta\textrm{V}$, respectively. 
The magnitude of the absorbed power enhances by increasing the thickness of the YIG layer. This is due to the fact that $P_{\texttt{abs}}$ is function of the volume of YIG ($\nu$), interacting with the microwave field ($h_{\texttt{rf}}$) following the equation: $P_{\texttt{abs}}=\pi\mu_{0}\nu f_{\texttt{FMR}}\chi^{\shortparallel}h_{\texttt{rf}}^{2}$. Here, $\mu_{0}$ and $\chi^{\shortparallel}$ are the permeability constant of vacuum and the imaginary part of the magnetic dynamic susceptibility, respectively. The general trend of $P_{\texttt{abs}}$ as function of the frequency is almost the same for the different thicknesses of YIG and presents two regimes. For a frequency lower than 3.3 GHz, the absorbed power is reduced. Higher than this frequency, $P_{\texttt{abs}}$ presents a nearly constant value around 0.4 and 10.0 mW for a thickness of 0.2 and 1 (also 3) $\mu$m, respectively. The reduction at low frequency of $P_{\texttt{abs}}$ is due to the reduction of the magnetic susceptibility\citep{Kurebayashi2011nmat}.

The YIG thickness dependence of the magnitude of $\Delta\textrm{V}$, as is shown in Fig.\ref{fig:Fig2}  b), does not present the same behaviour as $P_{\texttt{abs}}$. Decreasing the thickness of the YIG layer causes an increase of the magnitude of $\Delta\textrm{V}$. For example around 9 GHz, $\Delta\textrm{V}$ reaches 247, 28.5, and 14.5 $\mu$V at 10 mW for a YIG thickness of 0.2, 1, and 3 $\mu$m, respectively. Frequency dependences of $\Delta\textrm{V}$ for a YIG thickness of 1 and 3 $\mu$m present almost the same trend. $\Delta\textrm{V}$ increases in the frequency range of 1 to 2.4-2.8 GHz and up to 4 GHz the magnitude of $\Delta\textrm{V}$ is nearly constant. Note that the decrease of $\Delta\textrm{V}$ from the maximum to the constant value is abrupt which is not the case for the thinner YIG [0.2 $\mu$m]. One the other hand, $\Delta\textrm{V}$ obtained for this thickness [0.2 $\mu$m] is one of magnitude order larger than the value of $\Delta\textrm{V}$ for thicker YIG.

\begin{figure}
\centering
\includegraphics[width=9.3cm]{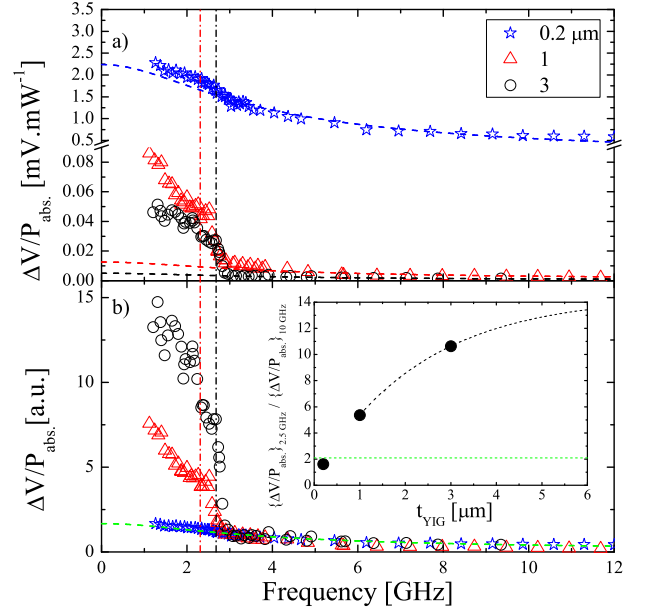}
\caption{\label{fig:Fig3} 
a) Frequency dependence of the conversion efficiency factor $\Delta\textrm{V}/P_{\texttt{abs}}$ for different thicknesses of YIG [0.2, 1, and 3 $\mu$m]. b) $\Delta\textrm{V}/P_{\texttt{abs}}$ normalized by the value at 3.3 GHz. The vertical dashed-dotted lines correspond to the frequency cutoff of the three-magnon splitting process estimated with the dispersion relation of spin waves for a YIG thickness of 1 (red) and 3 $\mu$m (black).
The green, blue, red, and black dashed curves correspond to the theoretical frequency dependence of $\Delta\textrm{V}/P_{\texttt{abs}}$ from Eq.\ref{DVPabs}. The inset in b) shows the dependence of the ratio of $\Delta\textrm{V}/P_{\texttt{abs}}$ at 2.5 and 10 GHz as function of the thickness of the YIG layer. All Measurements have been performed at room temperature under an rf excitation of 10 mW.
}
\end{figure}

In order to understand the frequency dependence of the spin current generation for the different thicknesses of YIG, we have calculated the factor $\Delta\textrm{V}/P_{\texttt{abs}}$ introduced by Kurebayashi \textit{et al.} \citep{Kurebayashi2011nmat} (see also Ref.\citep{SaitohFreqDep}) as is shown in Fig.\ref{fig:Fig3}. The normalization of $\Delta\textrm{V}$ by $P_{\texttt{abs}}$ confirms the existence of the abrupt change of $\Delta\textrm{V}$ demonstrated in Fig. \ref{fig:Fig2} b) which can be attributed to non-linear spin wave phenomena. This factor corresponds to the conversion efficiency of the angular momentum created by the microwave field into the spin current and it is described by the following equation extracted from Ref.\citep{Kurebayashi2011nmat}:
\begin{align}
   \begin{split}
       (a)~~~ & \dfrac{\Delta\textrm{V}}{P_{\texttt{abs}}}=A.\frac{1}{\sqrt{1+\left(  \dfrac{4\pi f}{\gamma M_{S}}\right)  ^{2}}} \\
       (b)~~~ & A=\dfrac{eL \Theta_{SH} g_{\uparrow\downarrow} \lambda tanh(t_\textrm{Pt}/2\lambda) }{\pi \mu_{0} \nu t_\textrm{Pt} \sigma M_{S}^{2} \alpha}\\
  \end{split} 
  		\label{DVPabs}
\end{align}
where $e$ is the elementary charge. $L$, $t_\textrm{Pt}$, $\lambda$, $\sigma$ are the distance between electrodes, the thickness of the Pt layer, the spin diffusion length, and the electric conductivity, respectively. $\Theta_{SH}$ and $g_{\uparrow\downarrow}$ are the spin-Hall angle and the spin mixing conductance. Parameters of the YIG layer are defined by the magnetization saturation $M_{S}$, the gyromagnetic ratio $\gamma$, and the Gilbert damping $\alpha$. Note that the prefactor $A$ presents a YIG thickness dependence introduced by $\nu$ which is the excited volume of YIG at the frequency $f$. 

The frequency dependence of the right part of Eq.\ref{DVPabs} (a) is introduced by the expression of the spin current\citep{PhysRevLett.88.117601,PhysRevB.67.140404}, $j_{s}$, and the magnetic susceptibility which can be written as:
\begin{equation}
\dfrac{\Delta\textrm{V}}{P_{\texttt{abs}}}\propto \frac{j_{s}}{f\chi^{\shortparallel}h_{\texttt{rf}}^{2}}\propto \int_0^{1/f}  \dfrac{1}{\chi^{\shortparallel}h_{\texttt{rf}}^{2}} \left\langle M(t)\times \dfrac{dM(t)}{dt} \right\rangle_{z} dt
\label{Integral}
\end{equation}
The right part of Eq.\ref{DVPabs} (a) is calculated by solving the Landau-Lifshiftz-Gilbert equation for the FMR condition of the integral in Eq.\ref{Integral} (see supplementary information in Ref.\citep{Kurebayashi2011nmat}). 


One can see in Fig.\ref{fig:Fig3} a) that for a frequency higher than 3.3 GHz, the experimental frequency dependence of $\Delta\textrm{V}/P_{\texttt{abs}}$ follows the theoretical behaviour calculated from Eq.\ref{DVPabs}. In the low frequency range (lower than 3.3 GHz), the evolution of this conversion factor ($\Delta\textrm{V}/P_{\texttt{abs}}$) is different between thinner [0.2 $\mu$m] and thicker YIG [1 and 3 $\mu$m] and several points should be made regarding this graph. First, for thicker YIG [1 and 3 $\mu$m], one can observed the same signature of the enhancement of $\Delta\textrm{V}/P_{\texttt{abs}}$ as demonstrated by Kurebayashi \textit{et al.} \citep{Kurebayashi2011nmat} for a YIG thickness of 5.1 $\mu$m. Nevertheless, despite the huge conversion factor of the thinner YIG (1.79 and 0.55 mV/mW$^{-1}$ at 2.5 and 10 GHz, respectively), no enhancement of $\Delta\textrm{V}/P_{\texttt{abs}}$ has been seen in the low frequency range [1-3.3 GHz] relative to the theoretical behaviour.

The good agreement between experiments and theoretical dependences from Eq.\ref{DVPabs} in the frequency range [3.3-12 GHz] confirms the fact that in this range $V_{\textrm{ISHE}}$ is directly proportional to the spin current generated by the magnetization precession of the uniform mode (long wavelength). Nevertheless, the enhancement of $\Delta V$ observed at low frequency for thicker YIG proves that a non-linear effect is present and means that the system absorbs the angular momentum from another source than the microwave field\citep{Kurebayashi2011nmat}. Because $V_{\textrm{ISHE}}$ is insensitive to the spin waves wavelength\citep{Kurebayashi2011nmat,2mag}, $\Delta V$ is not only defined by the uniform mode but from secondary spin wave modes, which present short-wavelength.

In Fig.\ref{fig:Fig3} b), $\Delta\textrm{V}/P_{\texttt{abs}}$ have been normalized by the values of this quantity at $f$=3.3 GHz for the different thickness of YIG. At low frequency, the normalized conversion efficiency is enhanced by increasing the YIG thickness. This observation is well represented in the inset of Fig.\ref{fig:Fig3} b), which shows the evolution of the ratio of $\Delta\textrm{V}/P_{\texttt{abs}}$ at 2.5 and 10 GHz as function of the YIG thickness. The enhancement of the YIG thickness form 1 to 3 $\mu$m induces an increase of this ratio from 5 to 11. For the thinner YIG, this ratio is equal to 2 and corresponds to the theoretical value represented by the horizontal green dashed line.

\begin{figure}
\centering
\includegraphics[width=10.5cm]{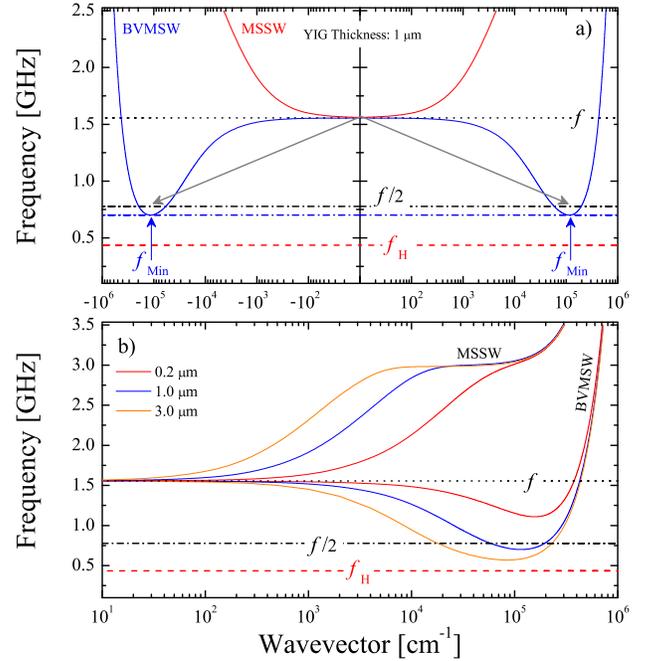}
\caption{\label{fig:Fig4} 
Dispersion relation of spin waves\citep{SWdispExchange,SWdispExchange2}. a) Dependence of the frequency, $f$, as function of the wavevector, $k$, when $k\parallel H$ (BVMSW) and $k\perp H$ (MSSW) for a YIG thickness of 1 $\mu$m. The arrows represent the three -magnon splitting process that creates two spin waves with a frequency corresponding to $f/2$. b) Evolution of the spin wave spectrum for different thicknesses of YIG [0.2, 1, and 3 $\mu$m]. For a) and b), the magnetic field is fixed at 150 Oe ($f=f_{\texttt{FMR}}$=1.55 GHz). $f_{\texttt{min}}$ is close to the Larmor frequency ($f_{\texttt{H}}$ in red) for thick sample of YIG. The exchange stiffness has been fixed at D=2 $10^{-13}$Oe$^{-1}$m$ ^{2}$ (see Ref.\citep{RezendeBEC}).
}
\end{figure}

The possibility to control the spin current at the YIG/Pt interface by the three-magnon splitting process has been demonstrated in Ref.\citep{Kurebayashi2011nmat}. This process is a non-linear effect easily actuated at low rf power (few $\mu$W) due to the low damping parameter of YIG, which is two orders of magnitude smaller than Permalloy (Ni$ _{19} $Fe$ _{81} $). Figure \ref{fig:Fig4} a) represents the dependence of the frequency, $f$, as function of the wavevector, $k$, when $k\parallel H$ (BVMSW) and $k\perp H$ (MSSW) estimated from the dispersion relation of spin waves from Refs.\citep{SWdispExchange,SWdispExchange2}. The arrows represent the three-magnon splitting process that creates two spin waves (short wavelength with $k\sim 10^5$ cm$^{-1}$) from the uniform mode (long wavelength with $k\sim 0$ cm$^{-1}$). The created magnons have a frequency corresponding to  half of the excitation frequency ($f/2$). The minimum of the BVMSW dispersion curve corresponds to the frequency minimum, $f_{\texttt{min}}$, which results from the competition between the magnetic field dipole interaction and the exchange interaction. The existence of such a dispersion minimum has been predicted by Kalinikos and Slavin \citep{SWdispExchange2}. A simple rule for the frequency range selection of the three-magnon splitting process came out from this graph. The three-magnon splitting process is allowed when $f/2>f_{\texttt{min}}$ and means that this non linear effect presents a frequency cutoff, $f_{\texttt{Cutoff}}$.

Figure \ref{fig:Fig4} b) shows the evolution of the spin wave spectrum for different thicknesses of YIG [0.2, 1, and 3 $\mu$m] with a magnetic field fixed at 150 Oe ($f=f_{\texttt{FMR}}$=1.55 GHz). $f_{\texttt{min}}$ depends of the thickness of the YIG layer which induces an evolution of $f_{\texttt{cutoff}}$ until a critical thickness where the three-magnon splitting process  is no longer allowed\citep{YIGPt01,Chernyshev}. When the thickness of the YIG increases, $f_{\texttt{min}}$ is approaching to the Larmor frequency $f_{\texttt{H}}$. In this particular case, $f_{\texttt{cutoff}}=\dfrac{2}{3}\gamma\mu_{0} M_{S}$ which is around 3.3 GHz for typical values of $\gamma$ and $M_{S}$. 

The vertical dashed-dotted lines presented in Fig.\ref{fig:Fig3} (and also Fig.\ref{fig:Fig2}) correspond to the frequency cutoff of the three-magnon splitting process estimated with the dispersion relation of spin waves\citep{SWdispExchange,SWdispExchange2} for a YIG thickness of 1 and 3 $\mu$m. For each thickness, $f_{\texttt{min}}$ have been calculated for different values of the magnetic field and a fixed exchange stiffness D=2 $10^{-13}$Oe$^{-1}$m$ ^{2}$ (see Ref.\citep{RezendeBEC}). We have observed that the three-magnon splitting process cease to exist  for the thinner YIG of 0.2 $\mu$m (thickness smaller than the exchange interaction length in YIG). At this specific thickness, the following 3-magnons splitting conditions $f/2<f_{\texttt{min}}$ is never reached for the whole frequency range. The theoretical values of the frequency cutoff for the YIG thickness of  1 and 3 $\mu$m are in good agreement with the experiments threshold, which are 2.35 and 2.7 GHz, respectively. 

Despite the fact that the three-magnon splitting is not allowed for the thinner YIG, the conversion efficiency of this sample presented in Fig.\ref{fig:Fig3} a) is huge compare to thicker YIG layer. In order to understand this enhancement, let's try to explain the factor $A$ from Eq.\ref{DVPabs}. This factor is defined by a product of fundamental constants ($e, \pi, \mu_{0}$), geometrical and materials parameters of YIG ($\nu, \alpha, M_{S}$), of Pt ($L, t_\textrm{Pt}, \lambda, \sigma, \Theta_{SH}$), and of the interface ($g_{\uparrow\downarrow}$). Tashiro \textit{et al.}\citep{YIGThickDep} have demonstrated experimentally that $g_{\uparrow\downarrow}$ is independent of the YIG thickness, which is consistent with the fact that the spin pumping (in the linear regime) is defined by the exchange interaction at the YIG/Pt interface. The magnetization saturation has been measured by vibrating sample magnetometer and pointed that this parameter is also independent of the YIG thickness (1760 G). $\gamma$ presents almost the same value for the different thickness of YIG which is between 1.81 and 1.82 $10^{7}$ Oe$^{-1}$rad.s$^{-1}$. This parameter has been extracted from the fitting of the experimental magnetic field dependence of the resonant frequency, $ f_{\texttt{FMR}} $, by using the Kittel equation\citep{Kittel}. In addition, parameters of the Pt layer are the same for the measured set of samples and the different deposition steps (Pt, Ti/Au, and Al$_{2}$O$_{3}$) have been done in the same time. Therefore, the only parameters in $A$ which can explain the observed large conversion efficiency for the thinner YIG are the YIG thickness and the damping parameter $\alpha$. 

Due to the fact that several modes contribute to the spin pumping (volume and surface), the FMR and the $V_{\textrm{ISHE}}$ lines in Fig.\ref{fig:Fig1} are broadened and in this case, one can only estimate the linewidth $\Delta\omega$ of the full spectrum. The linewidth determined in this way does not necessarily match with the intrinsic FMR linewidth, $\Delta\omega_{\textrm{FMR}}$, of the uniform mode, but is proportional to it ($\alpha\propto\Delta\omega_{\textrm{FMR}} \propto\Delta\omega$). By normalizing the factor $A$ with the YIG thickness, $t_\textrm{YIG}$, and $\Delta\omega$ (proportional at $\alpha$), we still observed an enhancement of $A$ for the thinner YIG of 0.2 $\mu$m. This increase is 4 times bigger than for thicker YIG (1 and 3 $\mu$m) which is not in agreement with the expected constant value of $A$ after the normalization.

There are many non-linear phenomena which can induce the creation of spin waves with short-wavelength. First, the enhancement of $A$ can be explained by a non-linear phenomenon, so-called the two-magnon process. This effect is due to the scattering of magnons on impurities and surfaces of the film and can contribute to the enhancement of the spin current at the YIG/Pt interface\citep{2mag}. Second, it is well known that the distribution of precession amplitude of the MSSW across the film thickness is exponential,  with its maximum at the surface of the film\citep{PhysRevB.77.214411}. The  BVMSW are characterized by a harmonic distribution of the dynamic magnetization across the film thickness, and thus is small at the surface of the film. 
The dynamic magnetization of the MSSW is localized at the surface and the contribution of these waves to the dc voltage from spin pumping is higher than the contribution of the BVMSW\citep{HilleSelection}. It might be possible that the contribution of these waves to the dc voltage generation is changed for thin YIG due to the fact that a reduction of the YIG thickness induces an enhancement (reduction) of the delay times (group velocity) of the spin waves.

%

In summary, we have reported the frequency dependence of the spin current emission in a hybrid YIG/Pt [6 nm] system as function of the YIG thickness [0.2, 1, and 3 $\mu$m] actuated at the resonant condition over a large frequency range [1-12 GHz]. We have demonstrate the possibility to control the efficiency of the spin current conversion by changing the YIG thickness and we have observed the threshold frequency dependence of the three-magnon splitting process. We have experimentally brought the evidence of the non-existence of this non-linear effect for a thin layer of YIG [0.2 $\mu$m] which is smaller than the exchange interaction length. On the other hand, the huge conversion factor $\Delta\textrm{V}/P_{\texttt{abs}}$ for the thinner YIG (1.79 and 0.55 mV/mW$^{-1}$ at 2.5 and 10 GHz, respectively) originates from another non-linear phenomena is the YIG, presents a better interest for the realisation of YIG-based devices.


We would like to acknowledge B. Wolfs, M. de Roosz and J. G. Holstein for technical assistance. This work is part of the research program (Magnetic Insulator Spintronics) of the Foundation for Fundamental Research on Matter (FOM) and is supported by NanoNextNL, a micro and nanotechnology consortium of the Government of the Netherlands and 130 partners, by NanoLab NL and the Zernike Institute for Advanced Materials.

\bibliography{YIGPt}

\end{document}